\documentclass[letterpaper,12pt]{article}
\usepackage[utf8x]{inputenc}
\usepackage{epsf}
\usepackage{wrapfig}
\usepackage{float}
\usepackage[pdftex]{xcolor,graphicx}
\usepackage{amsfonts}
\usepackage{amsmath}	
\usepackage{amssymb}
\usepackage{amscd}
\usepackage{mathrsfs}
\usepackage{float}
\usepackage{subfig}
\usepackage{lipsum}
\usepackage{footnote}
\usepackage{indentfirst}
\usepackage{setspace}
\usepackage[title]{appendix}
\usepackage[nocompress]{cite}
\usepackage{pdflscape}
\usepackage{fancyhdr}
\usepackage{enumitem}
\usepackage{array}
\makeatletter
\newcommand\appendix@section[1]{%
\refstepcounter{section}%
\orig@section*{Appendix \@Alph\c@section: #1}%
\addcontentsline{toc}{section}{Appendix \@Alph\c@section: #1}%
}
\let\orig@section\section
\makeatother

\fancypagestyle{plain}{%
\fancyhead{} 

}

\usepackage[normalem]{ulem}

\title{Flow of Carreau Fluids Through Non-Uniform Pore throats}

\author{Hassan Fayed$^1$\footnote{Corresponding author: adj. assistant professor, email: hehady@vt.edu }, \\
\small{$^1$University of Science and Technology in Zewail City, Egypt}}

\usepackage{stdclsdv}

\usepackage{hyperref}
\makeindex

\usepackage{nomencl}
\makenomenclature
\begin{document}

\hypersetup{%
    pdfborder = {0 0 0}
}

\maketitle

\begin{center}
   Abstract
\end{center}

The creeping flow of a generalized Newtonian fluid in a non--uniform pore throat is investigated analytically. The analytical solution determines the flow regimes and the transition point from Newtonian to the power law flow regime. As an application of the new model, a regular lattice--based pore network model is constructed to simulate the flows of shear--thinning and shear--thickening fluids through porous media. The pore throats have convergent--divergent geometries.

\section{Introduction}
The flows of Carreau fluids through porous media are important to many industrial applications such as enhanced oil recovery, drug delivery in organic tissues and polymer composites. The viscosity of these fluids displays complex behavior and depends on the local shear strain rates. The generalized Newtonian fluids are classified into two main categories;  shear--thinning fluids where the viscosity decreases with the increasing shear strain rates and shear thickening fluids where the viscosity increases with the increasing strain rates \cite{ carreau_1972}.  Polymer solutions are examples for the shear thinning fluids \cite{lopez_2003} and the dense granular suspensions are typical examples for the shear thickening fluids \cite {barnes_1989}. 

Porous media such as sand packs, sandstones, foam rubber, bread, lungs and kidneys are composed of a solid matrix that contains a system of random sizes pore voids \cite{BearBachmat}. These pore voids are interconnected to each other by another set of smaller void spaces and their structure is very complex and heterogeneous \cite{Kharusi_2007}. Therefore, the calculation of the pressure drop due to the flow of a generalized Newtonian fluid is a challenging task due to the complex structure of the flow conduits and the variable viscosity of the fluid. To study the flow of a generalized Newtonian fluid through a porous medium, three main approaches are used; pore-scale numerical simulation, pore-network modeling and the macroscopic continuum modeling.  In numerical pore scale simulations, the flow field through the pore voids is resolved in an image based domain. A computer tomography scanning (CT-image) is used to construct the 3D image of the microstructure of a porous medium sample. The Numerical methods are employed to resolve the flow field within the pores such as Finite Volume (FV), Finite element, Lattice Boltzmann Method (LBM) and Smooth Particle Hydrodynamics (SPH) \cite{sullivan_2006,mehran_2008,tiziana_2013,apiano_2009}. In this approach, the Carreau viscosity model \cite {carreau_1972} is used to determine the viscosity as a function of the magnitude of the local strain rate tensor.  The Newtonian and the power law behaviors are observed at low and high pressure drop values, respectively \cite {Shahsavari_2015}. However, resolving the flow on the pore scale needs large computational resources.  The second approach is the pore-network model where a pore network is constructed to represent the microstructure of a porous medium. The large voids are considered as pore bodies and they are interconnected to each other by smaller void spaces which are the pore throats. The pore networks can be constructed from a CT-scan of a porous medium \cite{Kharusi_2007}. The pore network model has proved to be successful in simulating single phase flows in porous media \cite{lopez_2003} and it is computationally cheaper than the pore scale numerical simulations. The flow rate as a function of the local pressure drop within each pore throat is predicted analytically in a constant radii--pore throats for Newtonian power law flow fluids by using Hagen--Poiseuille equation \cite{tan_2008,vogel_2000,blunt_2001,lopez_2003,balhoff_2006,christian_2006, raoof_2010}. The transition point from Newtonian to power law region is determined iteratively as dominstarted by Lopez ~\cite{lopez_2003}.   Fayed et al. \cite {fayed_2015} used Carreau model  Hagen--Poiseuille equation to obtain a graphical closed form solution in a uniform pore throat. The closed form solution predicts the flow regimes and the transition from Newtonian to power law regime in a uniform pore throat. The third approach is to study the fluid flows through porous media on a macroscale. The macroscopic approach considers a porous medium as a continuum where the fluid mobility through the complex microstructure is considered by the permeability tensor. The Kozeny--Carman \cite{carmen_1937}, Ergun \cite{ergun_1949}, Schneebeli \cite{schnebli_1959} and Wu et al. \cite{wu_2008} models are examples of the macroscopic permeability models for the Newtonian fluid flows in porous media \cite{balhoff_2004}. Other macroscopic models have been developed to model the flows of the power law fluids through porous media such as the Shenoy \cite{Shenoy_1993} and Tang and Lu \cite{tang_2014} models. These models estimate the permeability of a porous medium using Hagen-Poiselle where the variable viscosity is determined from the power law viscosity model. All of the permeability models have been developed for either a Newtonian fluid or a power law fluid. Therefore, using the Newtonian or the power law permeability models solely to predict the flow rates of a generalized Newtonian fluid through a porous medium is weakening the predictability of these permeability models. 
The knowledge of the transition point between the Newtonian and the power law flow regimes allows for accurate predictions of the flow rate under certain pressure drop. Adding to that, the permeability models rely on an empirical constant to account for the tortuous structure of the flow conduits. This empirical constant is not universal and depends on the microstructure of each porous medium sample \cite{ergun_1949}. 

In the pore network models of Carreau fluid flows, the power law viscosity model is commonly used to determine the fluid viscosity because of its simplicity Lopez ~\cite{lopez_2003}. However, the power law model predicts unphysical values for the viscosity in the limits of zero and infinite shear rates  as studied by Cannella et al. \cite{cannella_1988} and Vogel and Pusch \cite{vogel_1981}. On the other side, Carreau viscosity model \cite{carreau_1972} can not be used explicitly to obtain analytical solutions to be used in the pore network models. To overcome the shortcomings of the power law, Lopez \cite{lopez_2004} used a truncated power law model. The truncation shear strain rate at which viscosity law changes from the Newtonian to power law is known iteratively that increase the computational cost. 

\begin{figure}[H]
\centering
 \includegraphics[width=0.4 \textwidth,clip=true,trim=10 10 10 10]{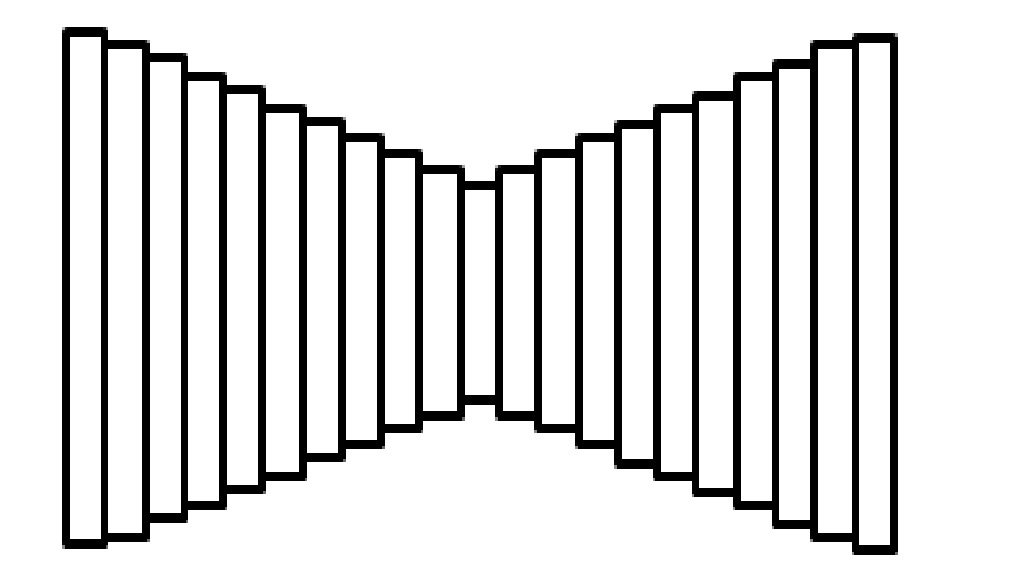}
 \caption{Schematic representation of the residual-based method}
 \label{fig:N0}
\end{figure}

The pore voids within a porous medium have a random size distribution and non--uniform pore geometry. The geometry of  pore--throats are approximated to have different cross sections \cite{blunt_2001}.  Sochi \cite{sochi_2015} developed a residual-based lubrication method to calculate the flow rate in a converging-diverging pore throat. This method is based on discretizing the fluid conduit into ring-like elements. Figure \ref{fig:N0} provide a graphical representation of this method.  
The analytical solutions of Ellis and Herschel-Bulkley fluids within each ring element is used to determine the local conductance of the flow.  However, this numerical treatment neglects the effects of the flow radial component on the pressure distribution along the pore throat. Therefore, a different treatment is needed to solve the flow in non--uniform pore throats where the effects of the radial flow velocity component on the pressure distribution are considered.  Also, Sochi \cite{sochi_2015} developed another semi--analytical model for Carreau fluid flows in uniform pore throats. This semi--analytical model require an iterative solver to determine the wall shear strain rate. 

 The paper adopts a non-dimensional procedure to simplify the governing equation to obtain an analytical solution for a creeping flow of generalized Newtonian fluid in a non--uniform flow conduit. Given a pore throat radius, the flow of a Carreau fluid displays a Newtonian and a power law behavior at low and high pressure drop values, Fayed et al. \cite{fayed_2015}. Therefore, to obtain an analytical solution for the flows of a Carreau fluid in a pore throat using the power law,  the flow is assumed to have a Newtonian behavior at low values of pressure drop and a power law behavior at high values to mimic the nature of the generalized Newtonian fluids. The transition from the Newtonian to the power law flow regimes takes place at certain value of the pressure drop in a specific. This pressure drop is termed here as the critical pressure drop. 

The main novelty of the paper is to develop an analytical solution for the transition pressure drop in a non--uniform pore throat. The advantages of the truncated analytical solution are demonstrated by implementing the solution in a pore network model where the model converges smoothly. The next section presents the derivations of the analytical solution and discusses the flow of a Carreau fluid in a converging--diverging pore throat. 
   
\section{Mathematical model}
The work in this paper is motivated by a recent study by Fayed et al.~\cite{fayed_2015} where the flow of a Carreau fluid displays a Newtonian and a power law behavior at low and high pressure drop values, respectively. Their observation confirms the theoretical analysis reported by Bourgeat and Mikelic ~\cite{mikelic_1996}. In this Section, the governing equations are presented and simplified to obtain an analytical solution for the transition pressure drop. 

Now, let us consider the flow of a Carreau fluid in a non--uniform pipe representing a pore throat. The flow is assumed to be steady, incompressible, axi--symmetric and isothermal where the viscosity is a function of the applied local shear strain rate. The variable radial profile of the pipe is given by $g^*(z^*)$ where $z^*$ is the axial coordinate. The pipe has an aspect ratio  $\displaystyle \delta=\frac{R}{L}$  where $R$ and $L$  are the pipe inlet radius and pipe length, respectively. The aspect ratio is assumed to be very small (i.e $\delta <<1$). The following non--dimensional parameters are adopted and will be used to simplify the governing equations: $z=z^*/L$, $r=r^*/R$, $g=g^*/R$, $\mu=\mu^*/\mu_o$, $Re= (\rho w_o R)/\mu_o$, $P=P^*R^2/(\mu_o w_o L)$, $u=u^*/(\delta w_o) $ and $w=w^*/w_o$ where $u$ and $w$ are the velocity components in the radial and axial directions respectively. $w_o$, $\rho$ and $\mu_o$ denotes the reference velocity, density and zero-shear-rate viscosity of the fluid, respectively.  $P$ is the static pressure and $Re$ is the Reynolds number. Now, using the Navier-Stokes equation in the cylindrical coordinates, the non--dimensional governing equations are,

\begin{equation}
\frac{1}{r}\frac{\partial (ru)}{\partial r}+\frac{\partial w}{\partial z}=0.
\label{eq:cont}
\end{equation}

\begin{eqnarray}\nonumber
Re \delta^3\left[u\frac{\partial u}{\partial r}+w\frac{\partial u}{\partial z}\right]&=& -\delta \frac{\partial P}{\partial r} +\delta^2\left[2\frac{\partial u}{\partial r}\frac{\partial \mu}{\partial r}+\delta^2 \frac{\partial u}{\partial z}\frac{\partial \mu}{\partial z}+\frac{\partial w}{\partial r}\frac{\partial \mu}{\partial z}+\right.\\
&& \left. \mu \left(\frac{1}{r}\frac{\partial}{\partial r}(r\frac{\partial u}{\partial r})\right)+\delta^2 \frac{\partial^2 w}{\partial z^2}  \right].
\label{eq:mom1}
\end{eqnarray}


\begin{eqnarray}\nonumber
Re \delta \left[u\frac{\partial w}{\partial r}+w\frac{\partial w}{\partial z}\right]&=&- \frac{\partial P}{\partial z} +\left[\frac{\partial w}{\partial r}\frac{\partial \mu}{\partial r}+\delta^2 \frac{\partial u}{\partial z}\frac{\partial \mu}{\partial r}+2\delta^2 \frac{\partial w}{\partial z}\frac{\partial \mu}{\partial z}+ \right. \\
&& \left. \mu \frac{\partial}{\partial r}\left(r\frac{\partial w}{\partial r}\right)+\delta^2 \frac{\partial^2 w}{\partial z^2}\right].
\label{eq:mom2}
\end{eqnarray}
 
A creeping flow is assumed where the Reynolds number is very small and on the order of $\delta$. In the frame work of the classical lubrication theory, the terms which are on the order of $\delta^2$ or higher can be neglected. As a result, the following set of governing equations are obtained as,

\begin{equation}
\frac{1}{r}\frac{\partial (ru)}{\partial r}+\frac{\partial w}{\partial z}=0.
\label{eq:conts}
\end{equation}
\begin{equation}
 \frac{\partial P}{\partial r} =0.
\label{eq:mom1s}
\end{equation}
\begin{equation}
\frac{\partial w}{\partial r}\frac{\partial \mu}{\partial r}+\mu \frac{\partial}{\partial r}\left(r\frac{\partial w}{\partial r}\right)=\frac{\partial P}{\partial z}. 
\label{eq:mom2s}
 \end{equation}      
In order to integrate Eq. \eqref{eq:mom2s} easily, it is written in a compact form as

\begin{equation}
\frac{1}{r}\frac{\partial}{\partial r}\left(\mu r\frac{\partial w}{\partial r}\right)=\frac{\partial P}{\partial z}. 
\label{eq:mom2ss}
 \end{equation} 
Eq. \eqref{eq:mom1s} shows that the pressure is uniform in the radial direction and varies only in the axial coordinate $z$ (i.e. P=P(z)). Eqs. \eqref{eq:mom2ss} and \eqref{eq:conts} are subject to the following boundary conditions

\begin{enumerate}[label=(\roman*)]
  \item at $r=0$, $\displaystyle\frac{\partial w}{\partial r}=0$ and $u=0$.
  \item at $r=g(z)$, $\displaystyle w=0$ and $u=0$.
  \item at $z=0$,  $p=P_i$ and $z=1$,  $p=P_o$.
\end{enumerate} 

The viscosity in Eq. \eqref{eq:mom2ss} is a function of the shear strain rate and can be defined by the Carreau viscosity model.  The viscosity profile of Carreau fluid consists of three main parts. The first part is the Newtonian plateau that describes the Newtonian nature of the fluid at low shear strain rates. The second part denotes for the power law behavior at intermediate values of the shear strain rates while the third part describes another Newtonian plateau at very high shear rate that is unlikely to occur as explained by Balhoff  \cite{balhoff_2005}. In this context, the Carreau viscosity model that describes the first Newtonian plateau and the power law region and given in the dimensional form as,

\begin{equation}
\mu=\mu_o   \left(1+(\lambda \dot{\gamma})^2   \right)^{\frac{n-1}{2}},
\label{eq:carre1}
\end{equation} 
where $\mu_o$ is the zero--shear--rate viscosity, $n$ is the power law index and $\lambda$ is the time constant. The power law index $n$ and the time constant $\lambda$ are parameters to be determined from the experimental viscosity profile of the fluid such as Xanthan gum solutions \cite{Escudier_2001}. At the critical shear strain, the transition from Newtonian plateau to the power-law region occurs. The critical strain rate determines the time constant $\lambda$ in the Carreau viscosity model. Despite the ability of the Carreau model to represent the first Newtonian Plateau, it is not possible to obtain an analytical solution on the pore scale. The power law model can be truncated to represent the viscosity profile of a Carreau fluid. In the literature, the truncation of the power law has been applied locally to overcome these two shortcomings of the power law model at the critical shear strain rate  (see \cite{lopez_2004} and  \cite{yao_2013}). However, the truncation process is not a straightforward and requires iterative procedure even in a simple straight pipe \cite{yao_2013}.  

For now, the power law viscosity model is used to calculate the viscosity in the momentum Eq. \eqref{eq:mom2ss} instead of the Carreau model.The Newtonian fluid is a special case of the power law at $n=1$. The dimensional form of the power law viscosity is given by

\begin{equation}
\mu^*=C\dot{\gamma}^{n-1},
\label{eq:dplaw}
\end{equation}
where $C$ and $n$ are the consistency coefficient and the power law index, respectively. These parameters are determined from the experimental viscosity profile. The dimensional local shear strain rate tensor for the assumed flow field in a single pore throat is written as,

\begin{equation}
	S=\begin{bmatrix} \left(\frac{\partial u}{\partial r}\right)^* & 0 &\left(\frac{\partial u}{\partial z}\right)^*\\ 0 & \left(\frac{u}{r}\right)^* & 0 \\\left(\frac{\partial w}{\partial r}\right)^*&0&\left(\frac{\partial w}{\partial z}\right)^*\end{bmatrix}. 
\label{eq:Smag}	
\end{equation}

By using the above nondimensional parameters and neglecting the terms which are on the order of $\delta^2$ or higher, the magnitude of the dimensional shear strain rate can be written as,

\begin{equation}
\dot{\gamma}=\frac{w_o}{R}\frac{\partial w}{\partial r}.
\label{eq:ssmag}
\end{equation}

From Eqs. \eqref{eq:dplaw} and \eqref{eq:ssmag}, the nondimensional form of the power law viscosity can be written as,
\begin{equation}
\mu=\frac{C}{\mu_o} \left(\frac{w_o}{R}\right)^{n-1}\gamma^{n-1},
\label{eq:dplawn}
\end{equation}
where $\displaystyle\gamma=\frac{\partial w}{\partial r}$. By integrating Eq. \eqref{eq:mom2ss} twice and apply the first boundary condition, an expression for the nondimensional axial velocity is obtained and given as,

\begin{equation}
w(r,z)=\frac{n}{n+1}\left(\frac{\mu_o}{2C}\right)^{\frac{1}{n}}\left(\frac{w_o}{R}\right)^{\frac{1-n}{n}}\left[r^{\frac{n+1}{n}}-g^{\frac{n+1}{n}}\right]\left(\frac{dp}{dz}\right)^{\frac{1}{n}},
\label{eq:w12}
\end{equation}
where, the axial velocity distribution is a function of the local pressure gradient $\displaystyle \frac{dp(z)}{dz}$ and the local pipe radius $g(z)$. From Eq. \eqref{eq:w12}, the derivative $\displaystyle \frac{\partial w}{\partial z}$ of the axial velocity is obtained and substituted into Eq. \eqref{eq:conts}. The resulting ODE equation is integrated and an expression for the radial velocity distribution is obtained as

\begin{eqnarray}\nonumber
u(r,z)&=&\frac{1}{n+1}\left(\frac{\mu_o}{2C}\right)^{\frac{1}{n}}\left(\frac{w_o}{R}\right)^{\frac{1-n}{n}}\left\{\frac{n+1}{2}\left(\frac{dp}{dz}\right)^{\frac{1}{n}}\frac{dg}{dz}g^{\frac{1}{n}}r- \right. \\
&& \left. \left(\frac{dp}{dz}\right)^{\frac{1-n}{n}}\frac{d^2 p}{dz^2}\left[\frac{n}{3n+1} r^{\frac{2n+1}{n}}-\frac{1}{2}g^{\frac{n+1}{n}}r \right] \right\}.
\label{eq:u1}
\end{eqnarray} 

Eqs. \eqref{eq:w12} and \eqref{eq:u1} can be rewritten for a Newtonian fluid where $n=1$ and $C=\mu_o$ as
\begin{equation}
w(r,z)=\frac{1}{4}\left[r^2-g^2\right]\left(\frac{dp}{dz}\right),
\label{eq:w1n}
\end{equation}
and
\begin{equation}
u(r,z)=\frac{1}{4}\left\{g\frac{dg}{dz}\frac{dp}{dz}-\frac{1}{4}\frac{d^2 p}{dz^2}\left[r^2-2g^2\right]\right\}.
\label{eq:u1n}
\end{equation} 

It is observed from these equations that the radial velocity distribution depends on the pressure gradient as well as the pressure Hessian $\displaystyle\frac{d^2 p}{dz^2}$. By using the boundary condition $u=0$ at $r=g(z)$ in Eq. \eqref{eq:u1}, an ODE for the pressure distribution along the axial coordinate $z$ is obtained and written as,

\begin{equation}
\frac{1}{3n+1}g\frac{d^2 p}{dz^2}+\frac{dg}{dz}\frac{dp}{dz}=0.
\label{eq:ddpz}
\end{equation}

The integration of Eq. \eqref{eq:ddpz} and using the third boundary condition, an analytical expression for the pressure distribution in the axial direction is obtained as 

\begin{equation}
p(z)=P_i-\frac{P_i-P_o}{I_p}\int_{0}^{z}{g(\zeta)^{-(3n+1)}}d\zeta,
\label{eq:pz}
\end{equation}
where
\begin{equation}
I_p=\int_{0}^{1}{g(\zeta)^{-(3n+1)}}d\zeta,
\label{eq:ipz}
\end{equation}
and the expression for the pressure gradient is given by,

\begin{equation}
\frac{dp}{dz}=\frac{-\Delta P}{I_p g^{(3n+1)}}.
\label{eq:dpz}
\end{equation}
The pressure distribution in the axial direction and the pressure gradient for a Newtonian fluids can be obtained by using $n=1$ as,

\begin{equation}
p(z)=P_i-\frac{P_i-P_o}{I_n}\int_{0}^{z}{g(\zeta)^{-4}}d\zeta,
\label{eq:pzn}
\end{equation}
where
\begin{equation}
I_n=\int_{0}^{1}{g(\zeta)^{-4}}d\zeta,
\label{eq:ipzn}
\end{equation}
and

\begin{equation}
\frac{dp}{dz}=\frac{-\Delta P}{I_n g^{4}}.
\label{eq:dpzn}
\end{equation}

Eqs. \eqref{eq:pz} through \eqref{eq:dpzn} show that the distributions of the pressure and its gradient in the axial direction depend on the geometry of the pore--throat as well as the fluid rheological properties of the fluid. On the pore scale, the flow changes its behavior from the Newtonian to the power law at certain value of the pressure drop \cite{fayed_2015}. The truncation of the power law in this paper is achieved analytically based on the flow average velocity to determine the critical pressure drop at which the flow changes its behavior in a non--uniform pore throat. First the average flow velocity in the pore throat is defined as,  

\begin{equation}
w_{av}=\frac{1}{\pi g^2(z)}\int_{0}^{g(z)}{2\pi r w(r,z)}dz.
\label{eq:int1}
\end{equation}

From this definition, the average velocity of a power law fluid is obtained and written as,
\begin{equation}
w_{av}=\frac{n}{3n+1}\left(\frac{\mu_o}{2C}\right)^{\frac{1}{n}}\left(\frac{w_o}{R}\right)^{\frac{1-n}{n}} \left(\frac{P_i-P_o}{I_p }\right)^{\frac{1}{n}} \frac{1}{g^2}.
\label{eq:wav1}
 \end{equation}
From Eq. \eqref{eq:wav1}, the average flow velocity depends on the local cross section of the pore throat which is a function of the axial coordinate $z$. To implement the truncation of the power law, it is important to obtain an expression for the average velocity of a Newtonian fluid by substituting $n=1$ as,

\begin{equation}
(w_{av})_N=\frac{P_i-P_o}{8I_n} \frac{1}{g^2}.
\label{eq:wav1n}
\end{equation}

At the transition point, the average flow velocity (i.e flow flux) from Eq. \eqref{eq:wav1} is equal to that from Eq. \eqref{eq:wav1n}.  Therefore, the critical pressure drop $\Delta P_{cr}$ is now determined from the intersection of these two equations and written in the dimensional form as,
\begin{equation}
\Delta p^*_{cr}=\frac{ L}{R}\left(\frac{8n\mu_o I_n}{3n+1}\right)^{\frac{n}{n-1}}\left(\frac{1}{2I_p C}\right)^{\frac{1}{n-1}}.
\label{eq:dpcr}
\end{equation}

It is important to note from Eq. \eqref{eq:dpcr} that the critical pressure drop depends on the inlet radius and the radial profile as well as the fluid rheological properties. These variables are constants in for a specific pore--throat. This means that the flow at any locality inside the pore throat is either Newtonian or follow power law behavior. A special case can be derived for a pore throat with a uniform radius $R$ where $g(z)=1.0$. Therefore, the critical pressure drop in the dimensional form for a uniform pore throat can be written as
\begin{equation}
\Delta p^*_{cr}=\frac{ L}{R}\left(\frac{8n\mu_o}{3n+1}\right)^{\frac{n}{n-1}}\left(\frac{1}{2 C}\right)^{\frac{1}{n-1}}.
\label{eq:dpcrc}
\end{equation}

\section{Case study of a Single Pore Throat}

\begin{figure}[H]
\centering
 \includegraphics[width=0.5 \textwidth,clip=true,trim=10 10 10 10]{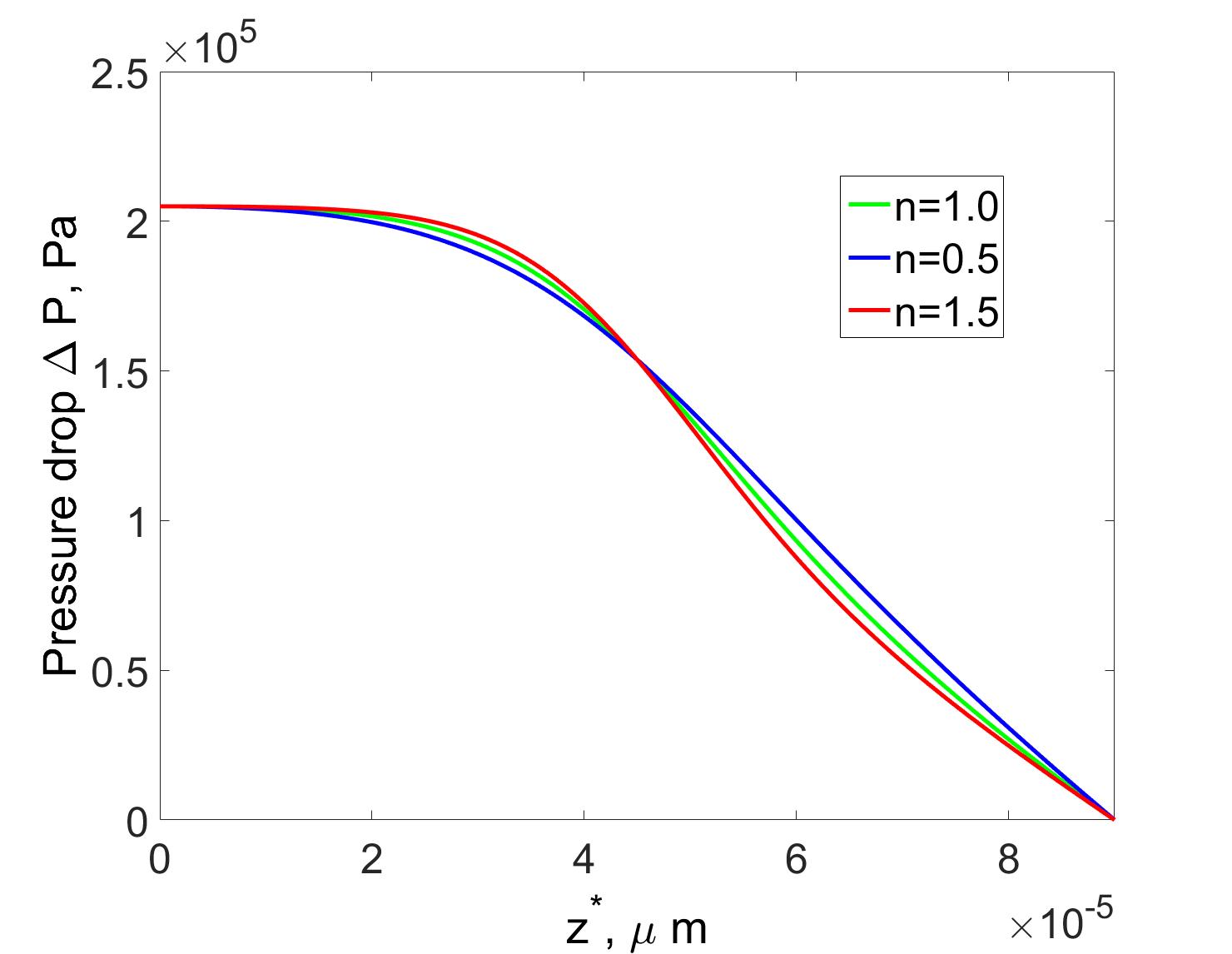}
 \caption{Pressure distribution in a converging-diverging pore throat for Newtonian $n=1$, shear thinning $n=0.5$ and shear thickening $n=1.5$ fluids at $\Delta P=2.05\times 10^5$}
 \label{fig:N2}
\end{figure}

The flows of  Newtonian,  shear thinning and shear thickening fluids have been solved in a single variable radius pore throat using the developed model in the paper. The nondimensional radial profile of the pore throat wall is described as,
\begin{equation}
g(z)=\frac{1}{2}+\frac{sin\left(\pi(z+1)\right)}{5},
\label{eq:pth1}
\end{equation}  
In this case study, the reference values used for the non-dimensionalization are $R=5\mu$m, $L=100\mu$m and $w_o$ is selected to be $10\mu$m/s. The Carreau fluid rheological properties are $\mu_o=1.0$ Pa.s and $\lambda=0.01$. The values of the consistency coefficient in the power law are determined graphically to be $C=9.77$ for $n=0.5$ and $C=0.1$ for $n=1.5$, so that the power law viscosity profile coincide with the Carreau profile at the intermediate shear strain rates. The value of the reference velocity $w_o$ is used in this Section to determine the dimensional value of the pressure drop through the pore throat. In the next section, the reference velocity will disappear when writing the expression for the flow flux in the dimensional form. The pressure profiles through the converging-diverging pore throat for the Newtonian, shear thinning and shear thickening fluids are depicted in Figure \ref{fig:N2}. 

\begin{figure}[H]
\centering
 \includegraphics[width=0.5 \textwidth,clip=true,trim=10 10 10 10]{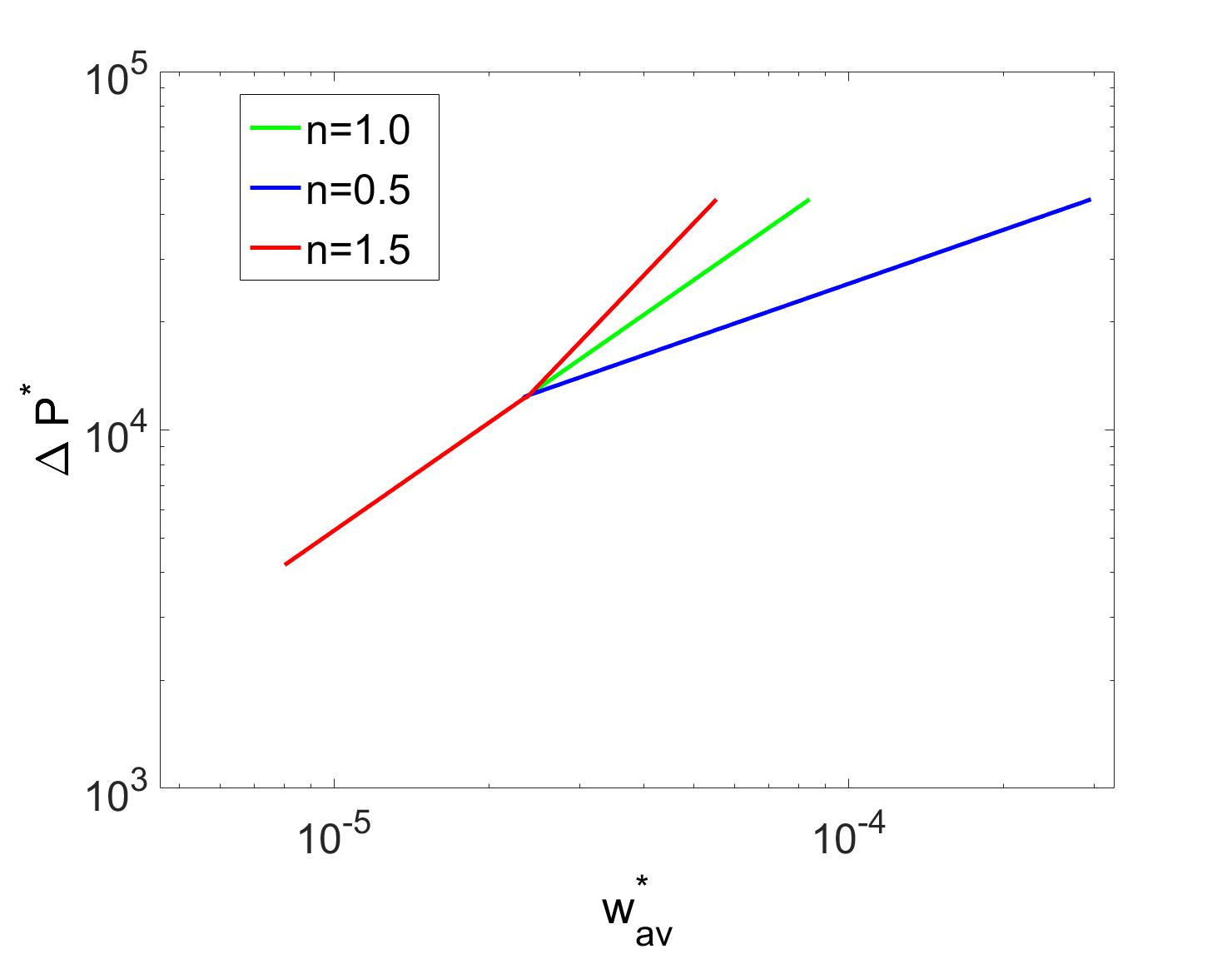}
 \caption{Pressure drop variations versus average flow velocity in a converging-diverging pore throat for Newtonian $n=1$, shear thinning $n=0.5$ and shear thickening $n=1.5$ fluids}
 \label{fig:N3}
\end{figure}

Let us look at the flow behavior in the given pore throat at different values of the pressure drop. The variations of the average flow velocity at the inlet section, $z=0$, with the pressure drop values have been monitored and reported at different values for the pressure drop as shown by Figure \ref{fig:N3} on a log-log scale. The flows of the shear thinning and shear thickening fluids are assumed to be Newtonian at pressure drop values lower than or equal to the corresponding critical pressure values, $\Delta P_{cr}=1.2487\times 10^4$ Pa for $n=0.5$ and $\Delta P_{cr}=1.2612\times 10^4$ Pa for $n=1.5$. The power law relationship given by Eq. \eqref{eq:wav1} is used at higher values of the pressure drop. On a log--log scale, the slop of the Newtonian curve is equal to unity while the slope of the power law curves is equal to the power law index.

As an application of the developed model to predict the macroscopic properties of the flows of generalized Newtonian fluids in porous media, next section presents a regular lattice--based pore network model. The pore throats are generated according a truncated log--normal distribution. The transition from the Newtonian to the power law behavior of the flow is investigated locally in each pore throat. The total average flow rate through the pore network is computed at each value of the total pressure drop and presented to study the macroscale flow behavior at different values of the porosity.

\section{Pore Network Generation}
There have been different approaches in the literature to generate pore networks to study transport phenomena through a porous medium. Some of these approaches are based on a regular Lattice-based pore network or scholastically generated pore networks \cite{raoof_2010b}. A CT-image have been also used to generate a more realistic pore network such as the pioneering work by Al-Kharusi and Blunt \cite{Kharusi_2007}.  Following Qin and Hassanizadeh \cite{qin_2015}, a 3D regular-lattice based network is generated to embody a representative elementary volume (REV) of a general porous medium sample. The pore bodies have a spherical shapes while the pore throats have a converging-diverging shapes. The spherical pore bodies are spaced at the lattice nodes and their sizes have been determined from a truncated log--normal distribution in the MATLAB code. The porosity of the sample can be changed through the mean pore size and the variance. The pore network side lengths $L_s$ is $2.0\times10^{-3}$m$\times$$2.0\times10^{-3}$m$\times$$2.0\times10^{-3}$m and the number of the pore bodies are $20^3$.

\begin{figure}[H]
\centering
 \includegraphics[width=0.8 \textwidth,clip=true,trim=10 10 10 10]{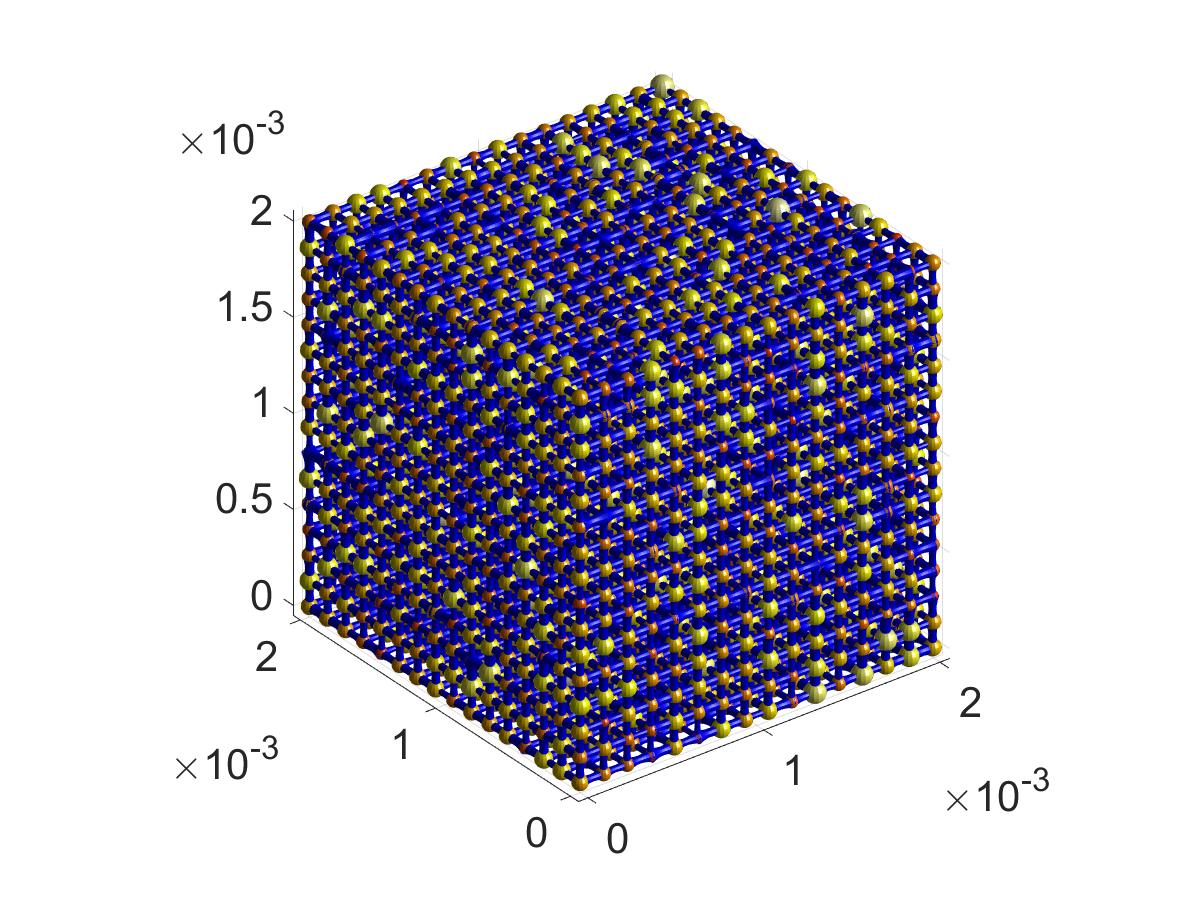}
 \caption{A regular lattice-based  pore network, $20\times 20\times 20$ pore bodies}
 \label{fig:N5}
\end{figure}

Each entire pore bodies are connected to $6$ adjacent pore bodies which are denoted by the coordination number. The coordination number of the boundary pore bodies are less than the entire ones and varies from $3$ at the corners to $4$ on the side edges and $5$ at the side faces as shown by Figure \ref{fig:N5}.  

\begin{figure}[H]
\centering
 \includegraphics[width=0.7 \textwidth,clip=true,trim=10 10 10 10]{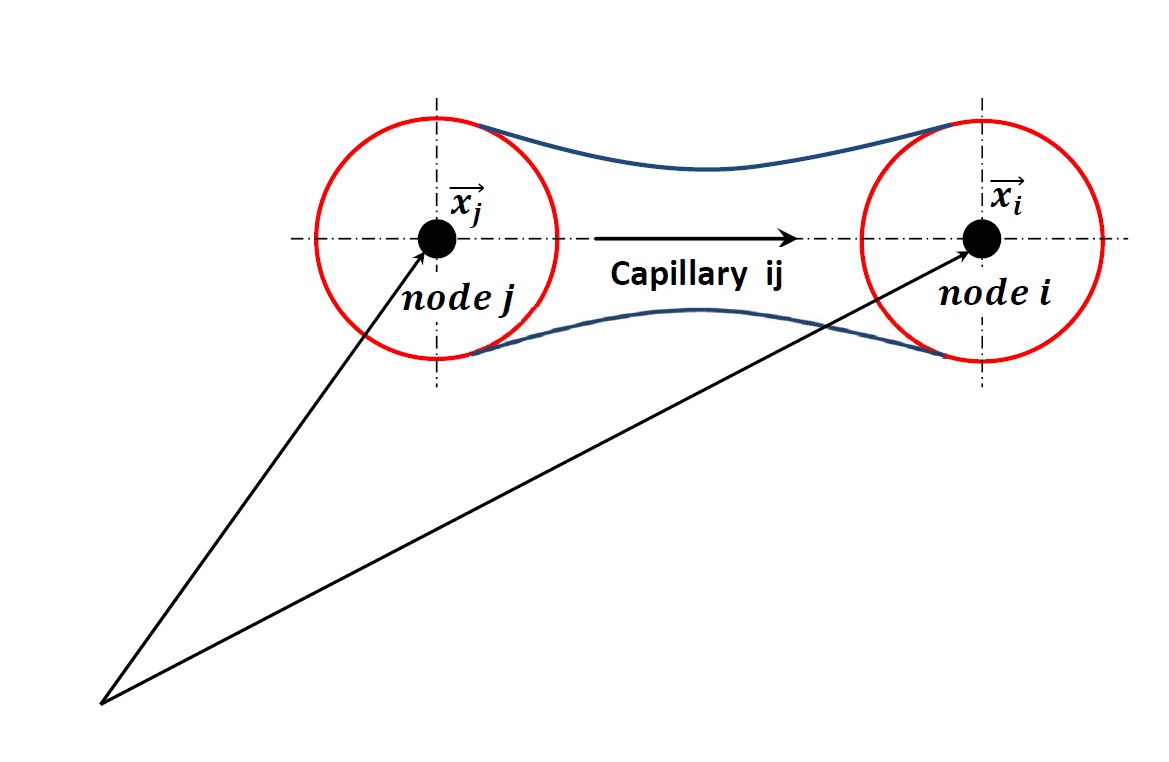}
 \caption{Schematic drawing of two interconnected pore bodies }
 \label{fig:N4}
\end{figure}

The pore throats that connect pore bodies are represented by a converging-diverging pipe similar to that given by Eq. \eqref{eq:pth1} where the inlet radius equal to $r_{ij}=min(r_i,r_j)$ and the length of each pore throat is approximated as $L=\sqrt{\Delta x_{ij}^2+\Delta y_{ij}^2+\Delta z_{ij}^2}-0.5(r_i+r_j)$. This approximation is shown by the schematic drawing in Figure \ref{fig:N4}.    
The porosity of the pore network equal the ratio of the total volume of pore void spaces to the total volume of the sample. For each sample, it is computed as $\displaystyle \epsilon =\frac{\sum V_{bodies}+\sum V_{throats}}{V_{sample}}$. In the next section, the pore network model and flow results are presented and discussed. 

\section{Pore Network Model}

\begin{figure}[H]
\centering
 \includegraphics[width=0.7 \textwidth,clip=true,trim=10 1 10 10]{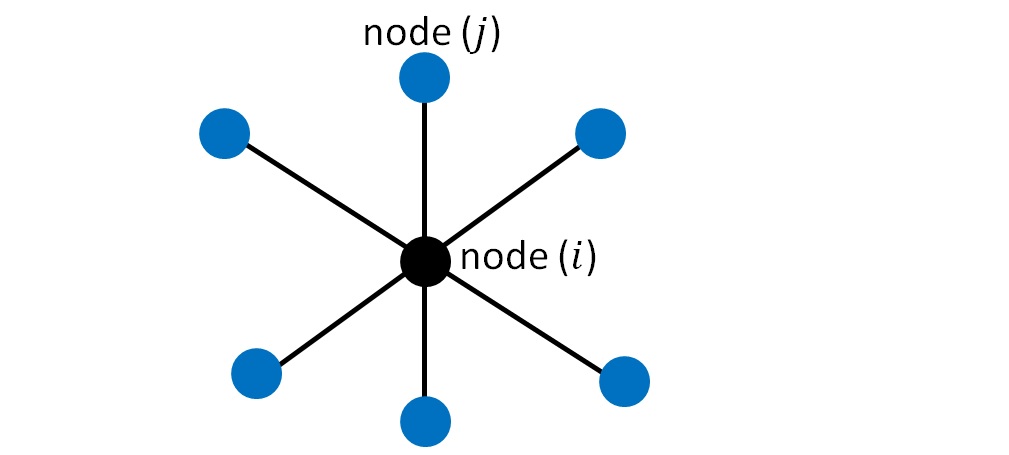}
 \caption{A schematic representation of interior pore bodies}
 \label{fig:N6}
\end{figure}

The flow of a Carreau fluid from a pore body $j$ to a pore body $i$ (see Figure \ref{fig:N6}) is assumed to be steady, incompressible and isothermal. This flow is driven by the pressure difference between each pairs of two connected pore bodies. The flow resistance is assumed to occur due to the skin friction on the walls of the pore throats (i.e. inertial terms are negligible) while the resistance in the pore bodies are negligible relative to that of the pore throats ~\cite{qin_2015}.  Eqs. \eqref{eq:wav1n} and \eqref{eq:wav1} provide the average flow velocity as a function of pressure drop between node ($i$) and its upstream node ($j$). At each node ($i$), the volume flow rate is conserved and the following equation is used

\begin{equation}
\sum_{j=1}^{N_j}{Q^*_{ij}}=0,
\label{eq:kij}
\end{equation}
where, $N_j$ is the coordination number of node $i$ and $Q_{ij}$ is the flow rate from node $j$ to node $i$. 

By using Eqs. \eqref{eq:wav1n} and \eqref{eq:wav1}, Eq. \eqref{eq:kij} can be written as

\begin{equation}
\sum_{j}^{N_j}{K^*_{ij}(p^*_j-p^*_i)}=0.
\label{eq:kij1}
\end{equation}
where $K^*_{ij}$ is the conductance of the pore throat that has variable radius. For the Newtonian flow region, the conductance is given by
\begin{equation}
K^*_{ij}=\frac{\pi}{8\mu_o}\left(\frac{ R^4}{ L I_n}\right)_{ij},
\label{eq:kij11}
\end{equation}
and $R_{ij}$ is the pore throat inlet radius at node $j$. For the power law flow region, the conductance is written as

\begin{equation}
K^*_{ij}= \pi \left(\frac{n}{3n+1}\right) \left(\frac{1}{2C}\right)^{\frac{1}{n}} \left[R^{\frac{3n+1}{n}}  \left(\frac{1}{I_p L}\right)^{\frac{1}{n}} \Delta p^{*\frac{1-n}{n}}\right]_{ij}.
\label{eq:kij12}
\end{equation}

Eq. \eqref{eq:kij1} is written at each pore body that results in large system of Algebraic equations and written as

\begin{equation}
\left[K\right]\left\{p^*\right\}=\left\{F\right\},
\label{eq:kij2}
\end{equation}
where the RHS of Eq. \eqref{eq:kij2}  comes from the known pressure at the boundary pore bodies at the right and left sides of the pore--network sample. The behavioral change from the Newtonian to the power law one depends on the local pressure difference between the pore bodies as well as the geometry of the pore throats.   Therefore, if the pressure difference between two adjacent pore bodies is less than or equal to the corresponding local critical pressure gradient, the conductance $K^*_{ij}$ will be calculated from Eq. \eqref{eq:kij11}. And the conductance of a specific pore throat $K^*_{ij}$ is calculated from Eq. \eqref{eq:kij12} if the pressure difference between two pore bodies is higher than the corresponding critical pressure drop value. This means that some elements of the matrix $\left[K\right]$ can be independent of the corresponding pressure drop $\Delta p^*$ and other elements are function of the local pressure drop in each pore throat which results in a non--linear problem that needs iterative solver. In the case of  Newtonian fluid flows, the pressure at the pore bodies is obtained from the solution of Eq. \eqref{eq:kij2} by direct inversion the matrix $\left[K\right]$. Lopez \cite{lopez_2003} and Sobrie et al. \cite{sorbie_2015} calculated the effective viscosity of a power law fluid in a pore network model iteratively using  Gaussian elimination method and determined the pressure drop at each pore body based on the effective viscosity in the corresponding pore throats. In this paper, the conductance of the pore throat for a Newtonian and/or a power law fluid is calculated directly from Eqs. \eqref{eq:kij11} and \eqref{eq:kij12} instead of computing the effective viscosity in each pore throat where the Newtonian and power law flow regimes in each pore--throat is determined analytically. This allows to simulate the flow of a Carreau fluid through a pore network effectively and predicting the transition from the linear Darcy flow regime to the non--linear power law regime on the macroscale for the given samples. A successive substitution iterative method has been used  to solve the system of equations defined by (\ref{eq:kij2}) by assuming that the flow is Newtonian as an initial guess. This iterative method converges very well in few number of iteration where the norm of $\displaystyle H(P)=\left[K\right]\left\{P^*\right\}-\left\{F\right\}$ becomes on the order of $10^{-13}$.   

\section{Results and Discussion}
The flow of a shear--thinning and shear--thickening fluids through a three representative pore networks has been studied at different flow rates through the pore network. The average flow velocity at the inlet pore bodies and the outlet pore bodies has been calculated to ensure the convergence of the iterative solver at each flow rate.   

\begin{figure}[H]
\centering
 \includegraphics[width=0.7 \textwidth,clip=true,trim=10 1 10 10]{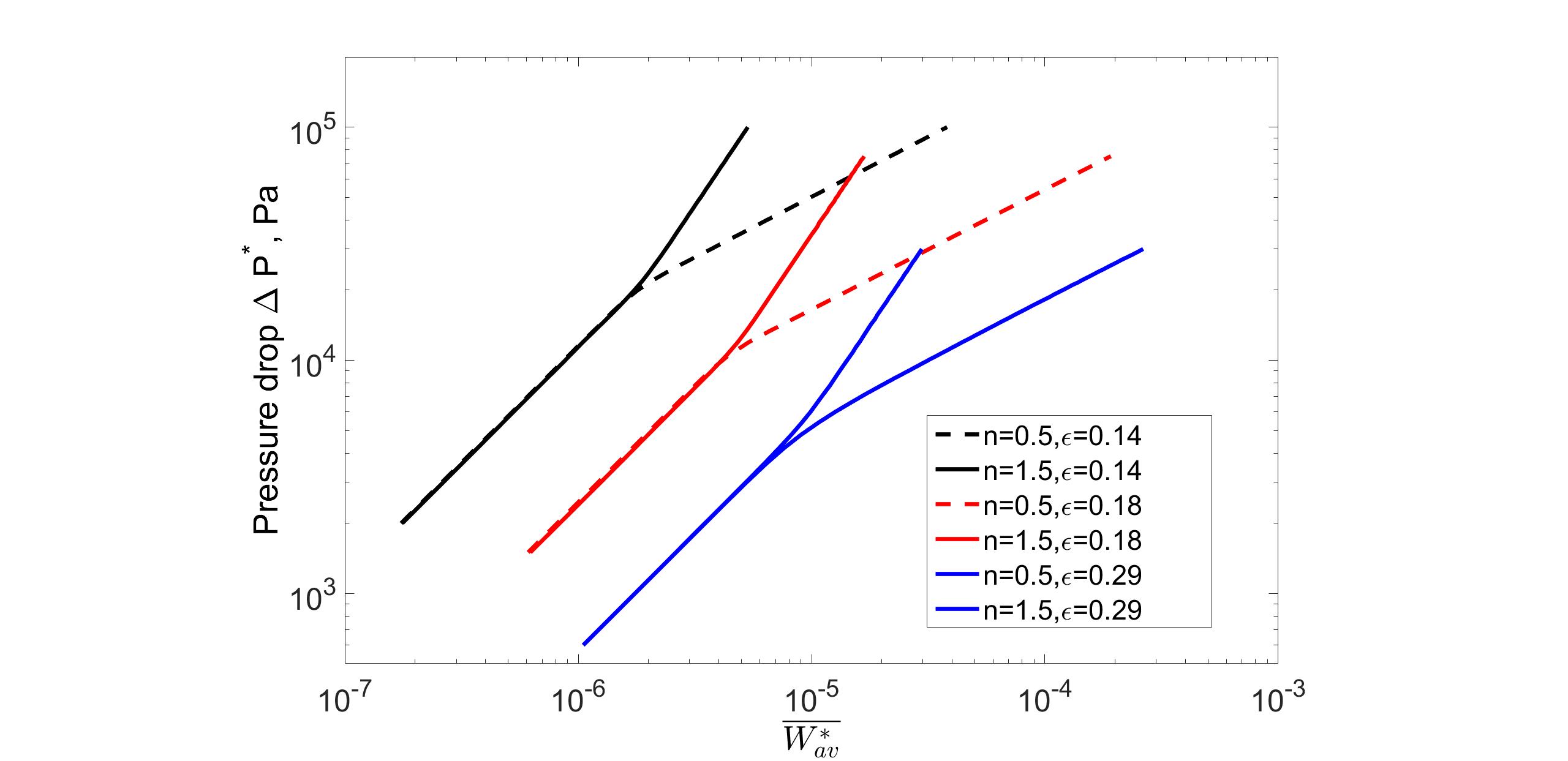}
 \caption{The total pressure drop variations with the Darcy velocity in pore networks representing different porous media for shear thinning and shear thickening fluids flows}
 \label{fig:N7}
\end{figure}

The conservation of volume flow rate at each pore body has been checked and the sum of the flow rates reaches to the zero--machine at each node. The pressure difference between the pore bodies at the inlet face and those at the outlet face (i.e total pressure drop) is monitored versus the flow superficial velocity (i.e Darcy velocity) at the inlet face and plotted on a log--log scale as shown by Figure \ref{fig:N7} for the three pore networks. 

The porosity of each representative sample is $\displaystyle \epsilon=0.14$, $\displaystyle \epsilon=0.18$ and $\displaystyle \epsilon=0.29$ for the sample1, sample2 and sample3, respectively. The other geometric parameters are listed in Table \ref{tab:gmppnt}. It is observed that flows of the shear thinning and shear thickening fluids in the pore network show a Darcy behavior at low values for the total pressure drop where the slop of the $\Delta P^*-\overline{W^*}_{av}$ curves is equal to the unity. This flow regime is named as linear Darcy region where the results are consistent with the Darcy equation given by \cite{balhoff_2005}, 
\begin{equation}
\frac{\Delta P^*}{L_s}=\frac{\mu_o}{K_D}\overline{W^*}_{av},
\label{eq:drcy1}
\end{equation}  
where, $\Delta P^*$ is the total pressure drop over a sample size of $L_s$ and $\overline{W^*}_{av}$ is the Darcy velocity,  $\mu_o$ denotes the fluid zero--shear rate viscosity.  $K_D$  is the Darcy permeability. The results from the pore network simulations are used to calculate the Darcy permeability by substitution in Eq. \eqref{eq:drcy1}. The permeability values of the three samples are calculated from Eq. \eqref{eq:drcy1} as given by the Table \ref{tab:knn}.

\begin{figure}[H]
\centering
 \includegraphics[width=0.7 \textwidth,clip=true,trim=10 1 10 10]{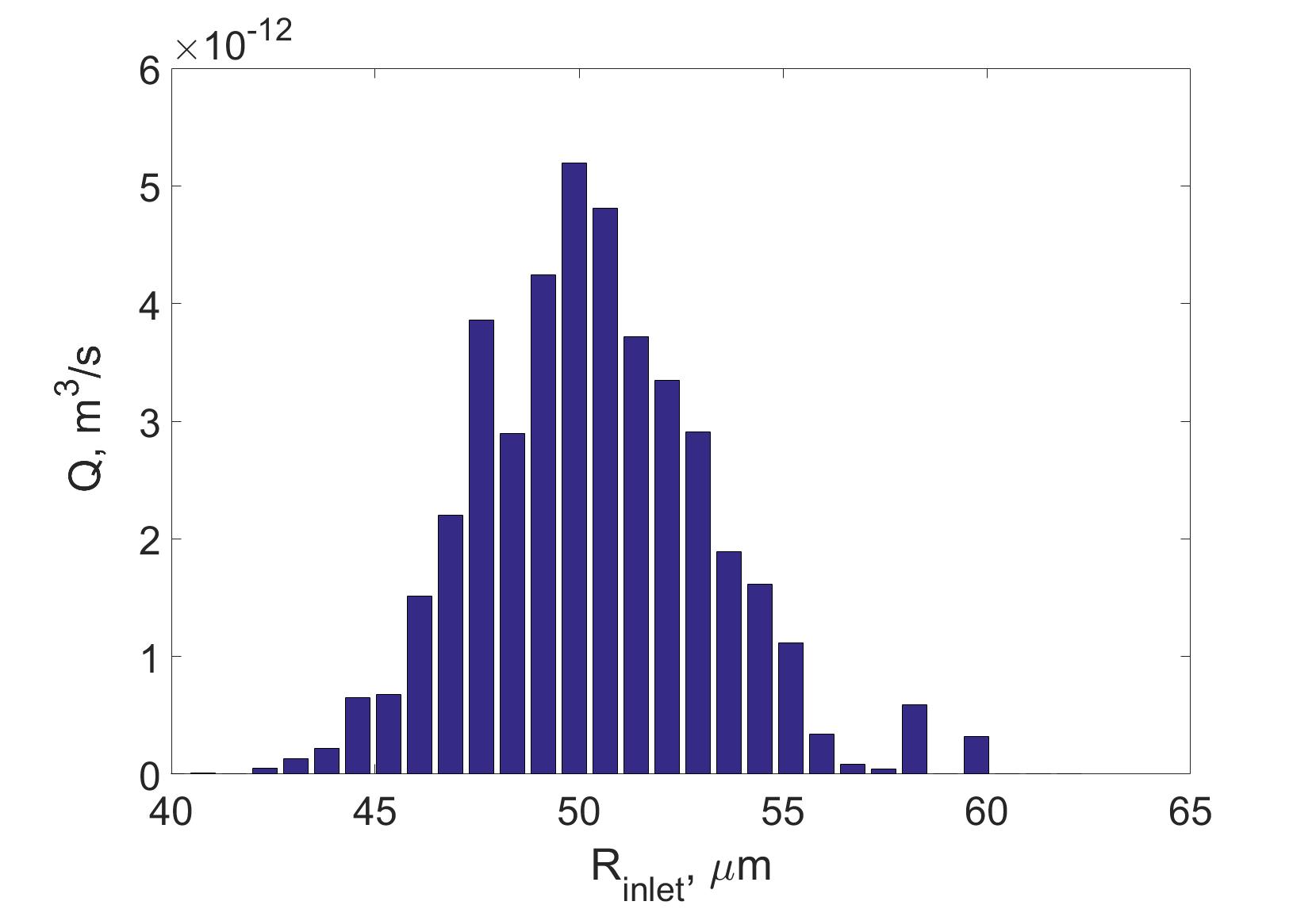}
 \caption{Histogram of flow rates in each pore throat inlet radius range in sample2 at $\Delta P^*=6000$ Pa; all pore throats experience Newtonian behavior}
 \label{fig:N71}
\end{figure}

When increasing the values of the total pressure drop, the slope of the ($\Delta P^*-\overline{W^*}_{av}$) curves change to be equal to the power law index which is 0.5 and 1.5 for the shear thinning and shear thickening fluids, respectively. The change in the slope takes place at the total critical pressure drop values given in Table \ref{tab:knn} for the three samples. Recalling that in a single pore throat, the transition from the Newtonian to the power law behavior is forced at local pressure drop values higher than the corresponding critical pressure drop where a sudden change in the slope of the ($\Delta p^*-w^*_{av}$) curves is discontinuous as shown by Figure \ref{fig:N3}. Despite the discontinuity in the slope on the pore scale, there is a smooth change in the slope on the macroscale and the slope of the macroscale curves is continuous (see figure \ref{fig:N7}). This means that the number of the pore throat that undergo transitional behavior from the Newtonian to the power law is changing gradually as the total pressure drop values are increased. To investigate the flow transition, a histogram of the flow rate versus the pore throat inlet radii is presented by Figure \ref{fig:N71}. 

It is important to observe that most of the flow rate goes through the medium size pore throats; very small flow rates go through the small and the large pore throats. this picture is totally different when using the bundle of parallel tubes models to simulate the flows of the Carreau fluids in porous media. In the bundle of tubes model, most of the flow rate will go through the largest capillary pipes while a small portion of the flow rates goes through the medium and small capillaries~\cite{sorbie_2015}. As shown in Figure \ref{fig:N71}, the medium pore throats in the neighborhood of 50 $\mu m$ are having most of the flow rates at $\Delta P^*=6000$Pa. Therefore, these pore--throats undergo the largest strain rates and the flow transition occur as a function of the medium pore--throat size. In the real porous media, the pore throats and pore--bodies have different irregular shapes and a random size distribution~\cite{Kharusi_2007}. Therefore, it is not feasible to calculate the integrals $I_p$ and $I_n$ for each pore throat in real samples. However, the medium pore--throat can be determined from CT-scan. An approach is proposed herein to predict the critical pressure drop on the macroscale, Eq. \eqref{eq:dpcr} is modified phenomenologically and given as, 

\begin{equation}
\left(\Delta P^*_{cr}\right)_{macro}=\chi \frac{L_s}{R_m}\left(\frac{8n\mu_o}{3n+1}\right)^{\frac{n}{n-1}}\left(\frac{1}{2 C}\right)^{\frac{1}{n-1}},
\label{eq:dpcr1}
\end{equation}
where, $R_m$ is the mean pore size. The effects of the pore irregular shapes and size distribution are considered through the constant $\chi$. Following from Eq. \eqref{eq:dpcr} the constant $\chi$ depends on the pore size distribution and the pore shapes. These two parameters are random and depend on the morphological structure of each porous medium sample. By comparing the predictions of Eq. \eqref{eq:dpcr1} and the values of the critical pressure drop listed in Table \ref{tab:knn}, the constant $\chi=$ 0.49, 0.44 and 0.21 for sample1, sample2 and sample3, respectively. At pressure drop values higher than the total critical pressure drop, the flow of the shear--thinning and shear--thickening fluids through the three networks displays power law behavior where the slope of the ($\Delta P^*-\overline{W^*}_{av}$) curves is equal to the power law index. This flow regime is named as the non--linear power law regime. It is so important to identify the linear Darcy and the non--linear power law regime when calculating the permeability of a certain porous medium. The distinction between the two regimes is achieved by the knowledge of the total critical pressure drop value.

The Darcy equation given by Eq. \eqref{eq:drcy1} is strictly developed for Newtonian flows in porous media. Therefore it is not valid in power law region. 
To obtain a Darcy-like relationship for the power law flow regimes, Eq. \eqref{eq:wav1} has been written in the dimensional form and simplified for a uniform pore throat radius. By using Carman's \cite{carmen_1937} geometric approximation, the mean hydraulic radius $\displaystyle R_h=\frac{\epsilon}{3(1-\epsilon)D_p}$ where $D_p$ is the mean grain size, the following Darcy-like relationship is obtained similar to those in the literature \cite{Christopher_1965} as, 
\begin{equation}
\frac{\Delta P^*}{L_s}= 2 C \left(\frac{3n+1}{n}\right)^n \left[\frac{3(1-\epsilon)}{D_p \epsilon}\right]^{n+1} (\overline{W^*}_{av})^n,
\label{eq:bkplaw}
\end{equation}

From Eq. \eqref{eq:bkplaw}, the Darcy--like equation is written as,
\begin{equation}
\frac{\Delta P^*}{L_s}= \frac{\mu_o}{K_p} (\overline{W^*}_{av})^n,
\label{eq:dlaw}
\end{equation}
where, $K_p$ is the power law permeability.
In this equation, Darcy--like equation, the pressure drop depends on the permeability, zero--shear rate viscosity and the average flow velocity raised to the power $n$. The difference between the three curves for Sample1, Sample2 and Sample3 is due to the differences in the permeability of each sample. From Eqs. \ref{eq:dlaw} and \eqref{eq:bkplaw} the permeability a porous medium is written as 

\begin{equation}
K_p= \tau \frac{\mu_o}{2C} \left( \frac{n\epsilon}{3n+1} \right)^n \left[ \frac{D_p \epsilon}{3(1-\epsilon)} \right]^{n+1},
\label{eq:dklaw1}
\end{equation}
where, $\tau$ is a constant that depends on the microstructure of a porous medium. In Eq. \ref{eq:dklaw1}, the ratio $\displaystyle\frac{\mu_o}{2C}$ is constant for a specific Carreau fluid. The above equation clearly shows that the apparent permeability in the power law region depends on the fluid properties such as consistency coefficient, zero--shear rate viscosity and the power law index. All the parameters in Eq. \eqref{eq:dklaw1} are known for a specific fluid flow in certain porous medium. The previous permeability models for the flows of the power law fluids in porous media are based on the effective viscosity.  The permeability of the power law region has been determined from the Eq. \eqref{eq:dlaw} using the results from the pore network simulations and given by Table \ref{tab:kpp}. It is observed that the permeability of the shear--thinning fluid in the power law region is three order of magnitude higher than that of the Newtonian flow region. While the permeability of the shear--thickening fluid in the power law region is three order of magnitude lower than that of the Newtonian region for sample1 and two order of magnitude lower than that of the Newtonian region for sample2 and sample3. 

In summary, the transition from the Newtonian to the power law regime of a Carreau fluid flows in a porous medium should be considered when calculating the macroscopic properties. The proposed model for the total critical pressure drop determines the transition point in terms of the mean pore size and the fluid rheological properties. The constant $\chi$, a correction constant, in this model accounts for the model uncertainty due to pores random size distribution and their irregular shapes.

\begin{table}[H]
	\centering
	\caption{Geometric parameters of the pore networks}
		\begin{tabular}{|p{3cm}|p{2cm}|p{2cm}|p{2cm}|}\hline
            Sample     & Mean pore radius&Pore radius variance&Porosity  \\ \hline
		        Sample1   & 30$\mu$m&20$\mu$m&0.14       \\ \hline
						Sample2   & 40$\mu$m&20$\mu$m&0.18 \\ \hline
						Sample3   & 50$\mu$m&30$\mu$m&0.29   \\\hline

		\end{tabular}
	
	\label{tab:gmppnt}
\end{table}
\begin{table}[H]
	\centering
	\caption{Permeability of the Darcy region and the total critical pressure drop}
		\begin{tabular}{|p{3cm}|p{3cm}|p{3cm}| }\hline
            Sample     & Permeability $K_D$ & $\Delta P^*_{cr}$\\\hline 
		        Sample1   & $1.75\times 10^{-13}$   &   $2.0\times 10^4$ \\\hline 
						Sample2   & $8.30\times 10^{-13}$ &     $1.35\times 10^4$\\\hline 
						Sample3   & $3.50\times 10^{-12}$  &    $0.5\times 10^4$\\\hline

		\end{tabular}
	
	\label{tab:knn}
\end{table} 
\begin{table}[H]
	\centering
	\caption{Permeability of the power law region, $K_p$}
		\begin{tabular}{|p{3cm}|p{3cm}|p{3cm}|}\hline
            Sample     & $n=0.5$ & $n=1.5$ \\\hline 
		        Sample1   & $1.30\times 10^{-10}$    & $2.36\times 10^{-16}$  \\\hline 
						Sample2   & $3.83\times 10^{-10}$    & $1.82\times 10^{-15}$ \\\hline 
						Sample3   & $1.15\times 10^{-09}$    & $1.05\times 10^{-14}$ \\\hline

		\end{tabular}
	
	\label{tab:kpp}
\end{table}

\section{Conclusions}
On the pore scale, the governing equations for the creeping flow of a Carreau fluid in a non--uniform pore throat have been simplified and solved analytically. The transition from Newtonian to the power law regime occurs at certain pressure drop that is defined as the critical pressure drop.  The transition point is found analytically as the intersection between the Newtonian and the power law flow equations for the average flow velocity.  The flow is modeled as Newtonian at pressure drop values lower than the critical pressure drop value whiles a power law fluid is considered at higher pressure drop values. 	
Regular lattice--based pore networks models are constructed as an application of the new model to study the flows of shear--thinning and shear--thickening fluids through porous media. The pore bodies have spherical shapes and their radii are obtained from a truncated log--normal distribution in the MATLAB code. The relationships for the average flow velocity are used to calculate the conductance of each pore throat in the Newtonian and the power law regimes. The pore network models produce a system of non--linear Algebraic equations  which have been solved iteratively. The variations of the Darcy velocity through the pore networks have been monitored and plotted versus the total pressure drop values. The flow on a macroscale displays a linear Darcy behavior at low values of total pressure drop and the non--linear power law behavior is observed at high values of the total pressure drop. The transition from the Darcy region to the power law region takes place at the certain pressure drop values termed as the total critical pressure drop. The values of total critical pressure drop have been determined from a semi-analytical model and compared with those values from pore-network results where good agreements were obtained.

\section{Conflict of Interest}
The author declares that there is no conflict of interest regarding the publication of this paper.

\bibliographystyle{ieeetr}
\clearpage
\bibliography{References}
\end{document}